\documentclass[aps,prl,twocolumn,a4paper,showpacs,dvips,floatfix]{revtex4}
\usepackage{graphicx}
\begin{document}

\title{Universality in the vibrational spectra of weakly-disordered two-dimensional clusters}

\author{\bf Gurpreet S. Matharoo}

\affiliation{Department of Physics, St. Francis Xavier University, Antigonish,
Nova Scotia B2G 2W5, Canada}
\date{November 17, 2008}

\begin{abstract}
We numerically investigate the vibrational spectra of single-component clusters in two-dimensions. 
Stable configurations of clusters at local energy minima are obtained, and for each the hessian matrix is 
evaluated and diagonalized to obtain eigenvalues as well as eigenvectors. We study the density of states so 
obtained as a function of the width of the potential well describing the two-body interaction.  As the width 
is reduced, as in three dimensions, we find that the density of states approaches a common form, but the two-peak 
behavior survives. Further, calculations of the participation ratio show that most states are extended, 
although a smaller fraction of the degrees of freedom are involved in these modes, compared to three dimensions.  
We show that the fluctuation properties of these modes converges to those of the Gaussian orthogonal ensemble 
of random matrices, in common with previous results on three dimensional amorphous clusters and molecular liquids.
\end{abstract}


\maketitle

\section{\bf Introduction}
The study of vibrational spectra has long provided important insights into the understanding
of amorphous states of matter~\cite{p01}, such as molecular glasses. However many 
key questions still remain unanswered. The vibrational disorder 
that is present in systems such as glasses is mainly topological 
in origin and over the years, a lot of effort has been put in to study glasses and
glass transitions in terms of underlying potential energy surface (PES)~\cite{p02,p03,p04,p05,p06,p07,p11,p12}.
One of the main focus has been the study of inherent structures, i.e local minima of PES and 
properties like aging have also been studied as a function of inherent structures~\cite{p13}.
Regarding vibrational spectra of systems having a topological 
disorder~\cite{p08,p09,p10,p14,p15,p16,p17,p20,p21,p22,p23,p23a,p24,p25,p26,fabian}, 
a popular approach rely on the study of statistical properties of random matrices, 
as within the harmonic approximation, the dynamical or the Hessian matrix contains 
all the dynamical features~\cite{p08,p09,p10,p14,p15,p16,p17}.

Recently, studies of vibrational spectra of amorphous systems have revealed 
some interesting universal aspects. In particular, studies of three-dimensional 
amorphous clusters~\cite{p19,p20,p21} have demonstrated the universality in the functional form of the 
density of states (DOS) and further, the statistical fluctuations have been shown to obey 
the characteristics of the Gaussian Orthogonal ensemble (GOE) of random matrices to a very 
high degree of accuracy. Recent studies on periodic, three-dimensional, molecular network-forming 
liquids have also shown that the statistical fluctuations obey the charactersistics of the
GOE~\cite{p23a} at various temperatures and for various densities. Even in amorphous alloys~\cite{fabian},
the statistical fluctutations have been reported to obey the characteristics of random matrix theory. 
Another universal feature that 
has been reported in earlier and recent studies on periodic three-dimensional amorphous systems 
using various model potentials is the density of states function approaching 
a limit that is independent of the explicit functional form of the potential in the
amorphous regime. The reasons for this universality have been suggested in~\cite{p18,p19,p20}.

The concepts of localization and delocalization of vibrational states has also been a focus in these studies,
with delocalized states being associated with GOE statistics~\cite{p16,p26} and localized states 
resulting in Poissonian statistics~\cite{p16}. A popular measure of localization is the 
participation ratio~\cite{p15,p16,p17,p20,p21,p22,p23,p23a,p24,p25,p26}. Values for participation 
ratios for three-dimensional amorphous clusters~\cite{p20,p21,p22,p23}, network-forming 
liquids~\cite{p23a} and atomic liquids~\cite{p26} indicate that the majority of the states are 
delocalized, supporting the apparent universality of GOE behavior in the statistical fluctuations.

The recent years have also seen an increase in the studies of vibrational properties of nanocrystalline materials
both theoretically~\cite{kara} and experimentally~\cite{fultz}. The new experimental techniques and methods 
have made it possible to explore the systems that are very small in size~\cite{fielicke} and increased 
computational power has even made it possible to explore such systems in reduced dimensions~\cite{hudon}. 

In the present work we numerically investigate the vibrational spectra of two-dimensional clusters. 
It is well known (e.g. in critical phenomena) that dimension has a strong effect on universal behaviors 
and a detailed study in a lower dimension would be a severe test of the robustness of the observed universalities 
in three-dimensional systems. E.g. as suggested in~\cite{p26a}, in quantum mechanical systems such 
as two-dimensional kicked rotors for lower dimensions $\left(d\leq 2\right)$, all the electronic 
states are exponentially localized, while for higher dimensions $\left(d > 2\right)$ there are extended 
as well as localized states. Also, disordered molecular systems display novel behavior in two-dimensions, 
e.g the occurance of the hexatic liquid phase that has a long range bond order without translational 
order~\cite{p26b}. Hence, checking the universality of the vibrational spectra for two-dimensional 
clusters is a severe test.  Our interest is to see the effect of dimensionality on the states 
being localized or delocalized with the main focus being on statistical fluctuations for the present 
two-dimensional case. 

\begin{figure}
\centerline{\includegraphics[width=3.0in]{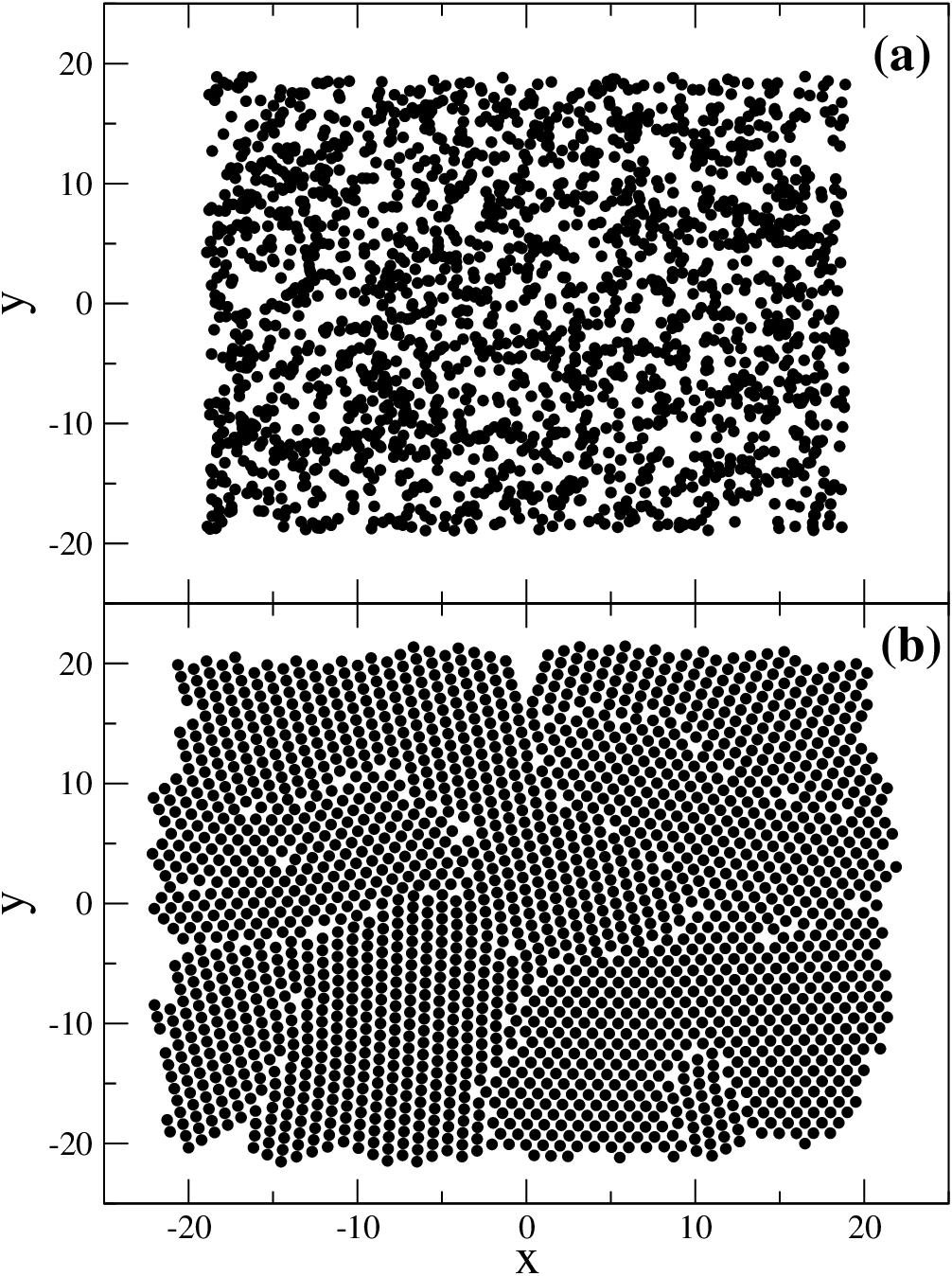}}
\caption{(a). A typical starting configuration of a two-dimensional cluster with $N = 2000$. 
(b). Stable two-dimensional cluster obtained using method of Homtopy with $N = 2000$.}
\label{fig1}
\end{figure}

\section {\bf Methodology and density of states}

The model used for the interactions between the particles is the Morse potential~\cite{p27}.
Over the years various studies have demonstrated the usefulness of this potential in 
simulations with references~\cite{p26c,charu} serving as a few examples. 
The form of the potential energy $V$ of the system is,
\begin{equation} 
V=\sum_{j>i}\{\exp\left[-2\alpha\left(r_{ij}-1\right)\right]-2\exp\left[-\alpha\left(r_{ij}-1\right)
\right]\}. 
\end{equation}
The potential is thus a function of inter-particle distance $r_{ij}$ and
the (positive) parameter $\alpha$; the sum is over all pairs of particles in the system. 
The $\alpha$ parameter can be tuned to fit a variety of systems ranging from
metals such as sodium with $\alpha = 3.15$; van der Waals bonded systems such as rare gases with
$\alpha =6$; to the very short-ranged interactions of the $C_{60}$ molecule with $\alpha = 13.62$~\cite{charu}. 
The values of $\alpha$ that have
been used in the present study are 3.5, 6.0, 10.0, 13.0 and 16.0. Using these values of $\alpha$ for the 
interaction potential, we first generate stable two-dimensional clusters with $N = 4000$, where $N$ is the 
number of particles in a cluster. For some calculations, clusters with $N = 500$ and $N = 2000$ particles are used. 

\begin{figure}
\centerline{\includegraphics[width=3.0in]{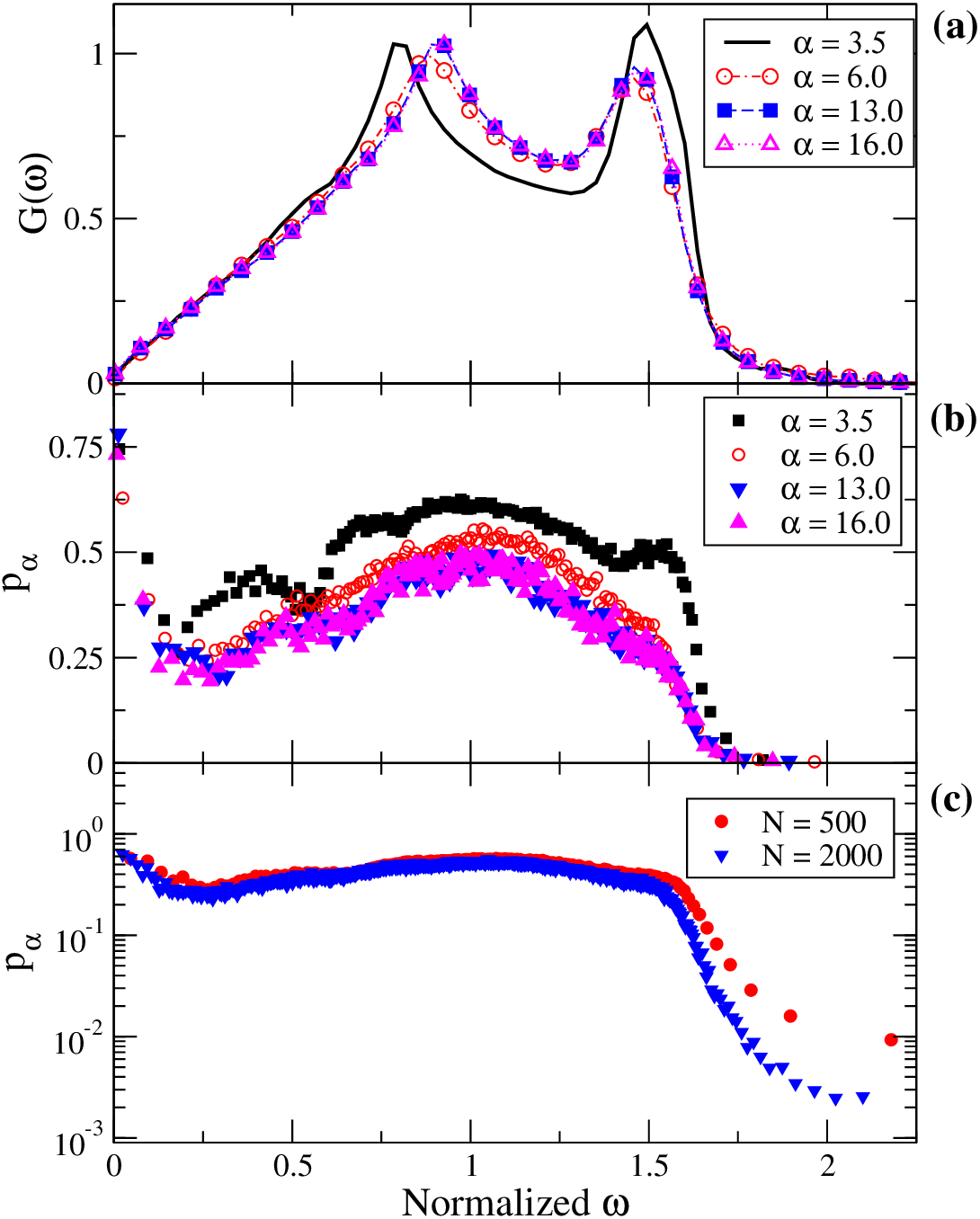}}
\caption{Color online (a). Density of states $\left(G\left(\omega\right)\right)$ vs. normalized frequencies of
normal modes plotted for various values of $\alpha$. Area under each curve has been rescaled to
unity. $N = 4000$ has been used in this calculation. (b). Participation ratios $\left(p_{\alpha}\right)$
vs. normalized frequencies of the normal modes for various values of $\alpha$. 
(c). $\log\left(p_{\alpha}\right)$ vs. normalized frequencies of the normal modes for $\alpha = 6.0$ for 
$N = 500$ and $N = 2000$ particles.}
\label{fig2}
\end{figure}
In order to generate stable clusters, we begin by initializing the
particle positions within the limiting distance $r = \sqrt{N/\pi}$ on the x-y plane. Fig. 1~(a) shows 
an example of a typical starting configuration for $N = 2000$. In order to obtain the potential energy minimum for
such a configuration, we use the homotopy method of minimization~\cite{p28,p29}. According to this method, in 
order to find a local minimum of a function $V$, we minimize in a series of steps the 
quantity $\theta V + \left(1-\theta\right)W$, where $W$ is a suitably chosen simple 
function (e.g. quadratic) and $\theta$ is varied from 0 to 1 in a finite number of steps, 
typically 20. The minimized configuration for one value of $\theta$ serves as the initial configuration 
for the subsequent step. Fig. 1~(b) illustrates a two-dimensional stable cluster generated by this method.   

At each local minimum generated by this method, a hessian matrix is constructed using the position 
coordinates of particles at that minimum and is diagonalized using standard methods to obtain the 
eigenvalues as well as the eigenvectors.  The eigenvectors correspond to the normal modes and 
the eigenvalues~$\left(\lambda\right)$ are related to the frequencies~$\left(\omega\right)$ 
of the obtained normal modes according to $\omega = \sqrt{\lambda}$. The normal modes obtained in this manner
are also known as quenched normal modes (QNM). A quantity that is of central physical interest is
the density of states (DOS) function, the histogram of values of $\omega$, denoted by G$\left(\omega\right)$. 
\begin{figure}
\centerline{\includegraphics[width=3.0in]{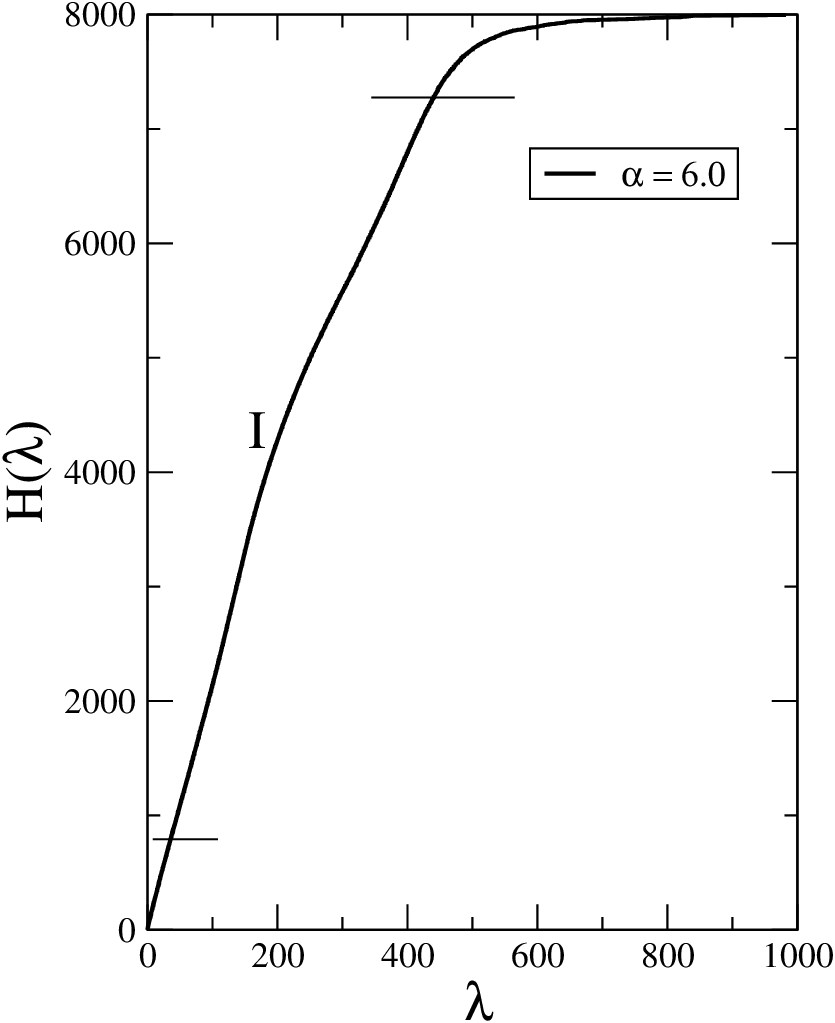}}
\caption{Integrated density of states for a single spectrum for $\alpha = 6.0$ with $N = 4000$.
Data in the region I has been used for unfolding and to analyze the statistical fluctuations}
\label{fig3}
\end{figure}
In Fig. 2(a), we plot $G\left(\omega\right)$ for several values of $\alpha$. To improve statistics we combine 
information from the available distinct quenched configurations (atleast 75 for each case) by averaging over all the 
configurations. After obtaining the raw averaged DOS function, the average frequency is normalized
to unity and G$\left(\omega\right)$ is rescaled so that the area under the curve is also unity. This process
enables us to compare the DOS for different values of $\alpha$. 
Fig. 2(a) clearly shows the existence of a two-peak behavior for all the values of $\alpha$ studied which
suggests that the obtained local minima have weakly-disordered domains with grain boundaries. This
is further evident from Fig. 1(b). This two-peak nature observed in the DOS function is also 
consistent with the recent results obtained by Hudon et al.~\cite{hudon} where bulk nanocrystalline materials 
in two-dimensions are studied using Lennard-Jones interaction potential and a similar form for $G(\omega)$ 
is observed and it also shows that the present system behaves as a ``normal'' two-dimensional 
system {\it as per} the notation used by Hudon et al~\cite{hudon}. The peaks shift towards right with 
increasing $\alpha$ and the overall structure varies very slowly for higher values of $\alpha$.  

This is also notably different than the earlier study conducted on a periodic three-dimensional system 
using a similar generalized Morse potential~\cite{p20,p23}. In the three-dimensional case, 
it was possible to generate states with variable amounts of disorder and the analysis of 
disordered states showed that the two-peak behavior of DOS changed over to a one-peak as 
the value of $\alpha$ was increased and by the time $\alpha$ reached 16, only one peak remained. 
Further increase in $\alpha$ values lead to small changes in the overall DOS shape~\cite{p20,p23}. However in the 
present case, potential energy minimization results in states that are weakly-disordered rather 
than being totally amorphous and at present, it is not clear how one could generate stable clusters 
in two-dimensions that have variable amounts of disorder. However a similar feature at both two and 
three dimensional case occurs at higher values of $\alpha$, where DOS function varies very slowly with 
further increases in $\alpha$.  

Fig. 2(b) shows the participation ratios~\cite{p15,p16,p17,p21,p22,p23,p23a,p24,p25,p26}
calculated using the eigenvectors corresponding to each eigenvalue for all values of $\alpha$.
Mathematically, the participation ratio is defined as,
\begin{equation} 
p_{\alpha}~\equiv~\left[N\sum_{i}\left(e_{\alpha}^{i}.e_{\alpha}^{i}\right)^{2}\right]^{-1} 
\end{equation}
where $e_{\alpha}^{i}$ is the projection of the eigenvector (labelled by $\alpha$) onto particle $i$. 
For extended modes, $p$ is of the order of unity and does not depend on system size, while for localized 
modes it will scale inversely with the system size. 
Calculations for participation ratios have been done using $N = 500$ and $N = 2000$ particles.
For each value of $\alpha$ on this plot, information from various quenched configurations is combined
to improve statistics. For all the $\alpha$ values studies, maximum values for participation ratios stay below
or close to 0.6 which might suggest that more states are localized in two dimensions as compared to three-dimensions.
However, closer examination of participation ratios plotted on a semi-log scale in Fig. 2(c) with varying $N$ shows
that the states are definitely not localized over the wide spectrum that has been used for studying statistical
fluctuations. Since the two systems differ in size by a factor of 4, localized modes in the larger system should 
have participation ratios approximately 1/4 of those in the smaller system. We find that differences on this 
scale only occur towards the higher values of $\omega$. Hence, even though the participation ratios in the
middle part of the frequency spectrum that is used in the analysis of statistical fluctuations are close to
0.5 or 0.6, these still have behavior consistent with extended modes of the system.

\section{Fluctuations}
  
We now investigate the statistical fluctuation properties of the DOS.  
In the present case, the fluctuation properties are computed for $\lambda$'s, the eigenvalues.  
For a particular inherent structure, we denote the elements of the obtained spectra
by $\lambda(i)$ with $i = 1,2,\ldots, 2N$. Since the present system is two-dimensional, the
first three elements in the spectra will be zero and the remaining $(2N -3)$ positive
frequencies are characterized by defining a mean local density as well as the fluctuations around it.
The first step in computing the fluctuation properties is to unfold the data~\cite{p21,p22,p23,p23a,p30,p31,p32}. 
This process enables us to transform the eigenvalues in such a way that the average spacing between two successive 
eigenvalues is unity.

\begin{figure}
\centerline{\includegraphics[width=3.0in]{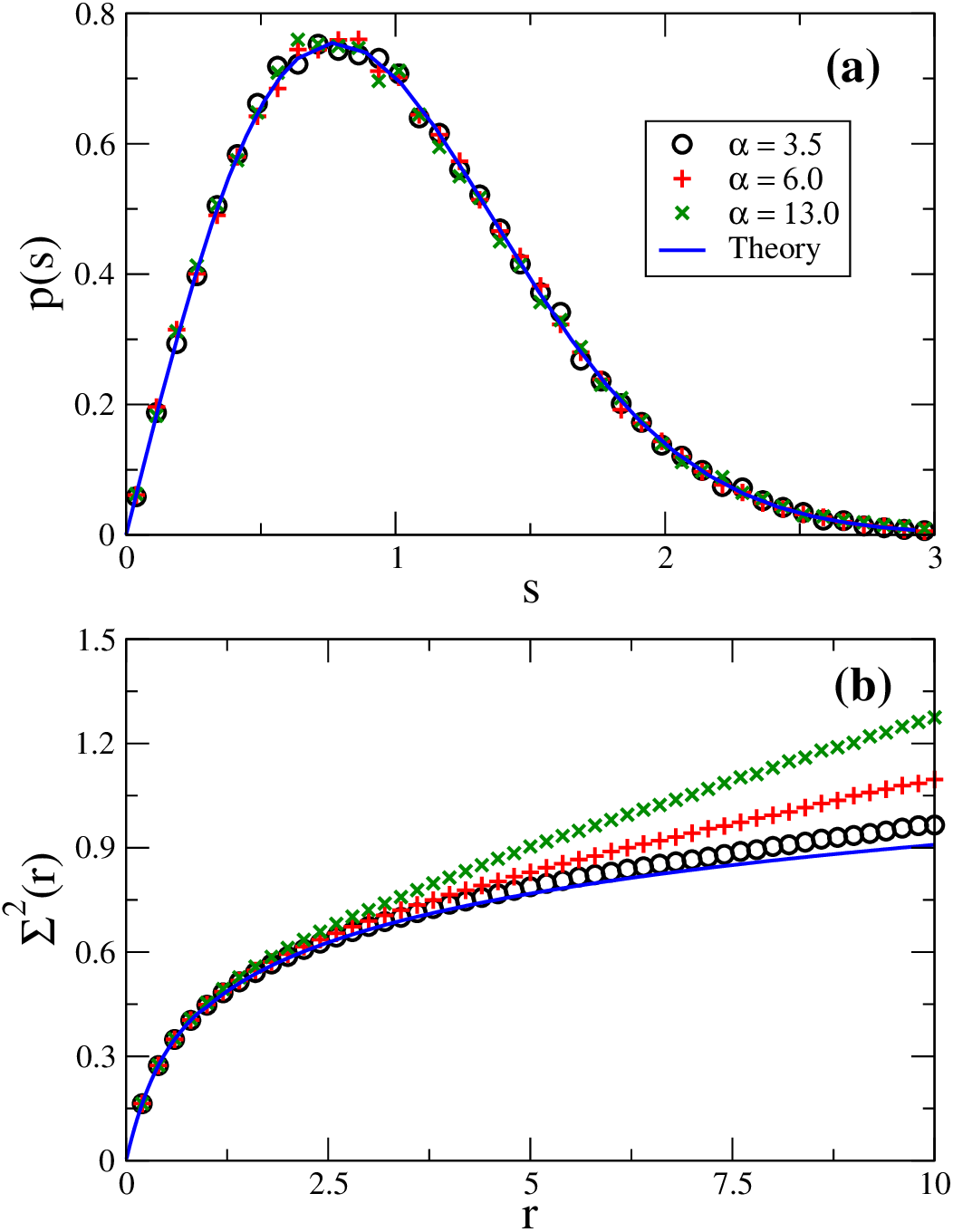}}
\caption{Color Online (a). Probability density $p\left(s\right)$ for normalized
nearest neighbor spacing~$\left(s\right)$ for various values of $\alpha$. Also shown is the
prediction for the GOE. $N = 4000$ has been used in this calculation.
(b). Variance of the number of levels in intervals of length $r$ shown as
a function of $r$ for various values of $\alpha$. Also shown is the prediction for the GOE.
$N = 4000$ has been used in this calculation.}
\label{fig4}
\end{figure}

\begin{figure}
\centerline{\includegraphics[width=3.0in]{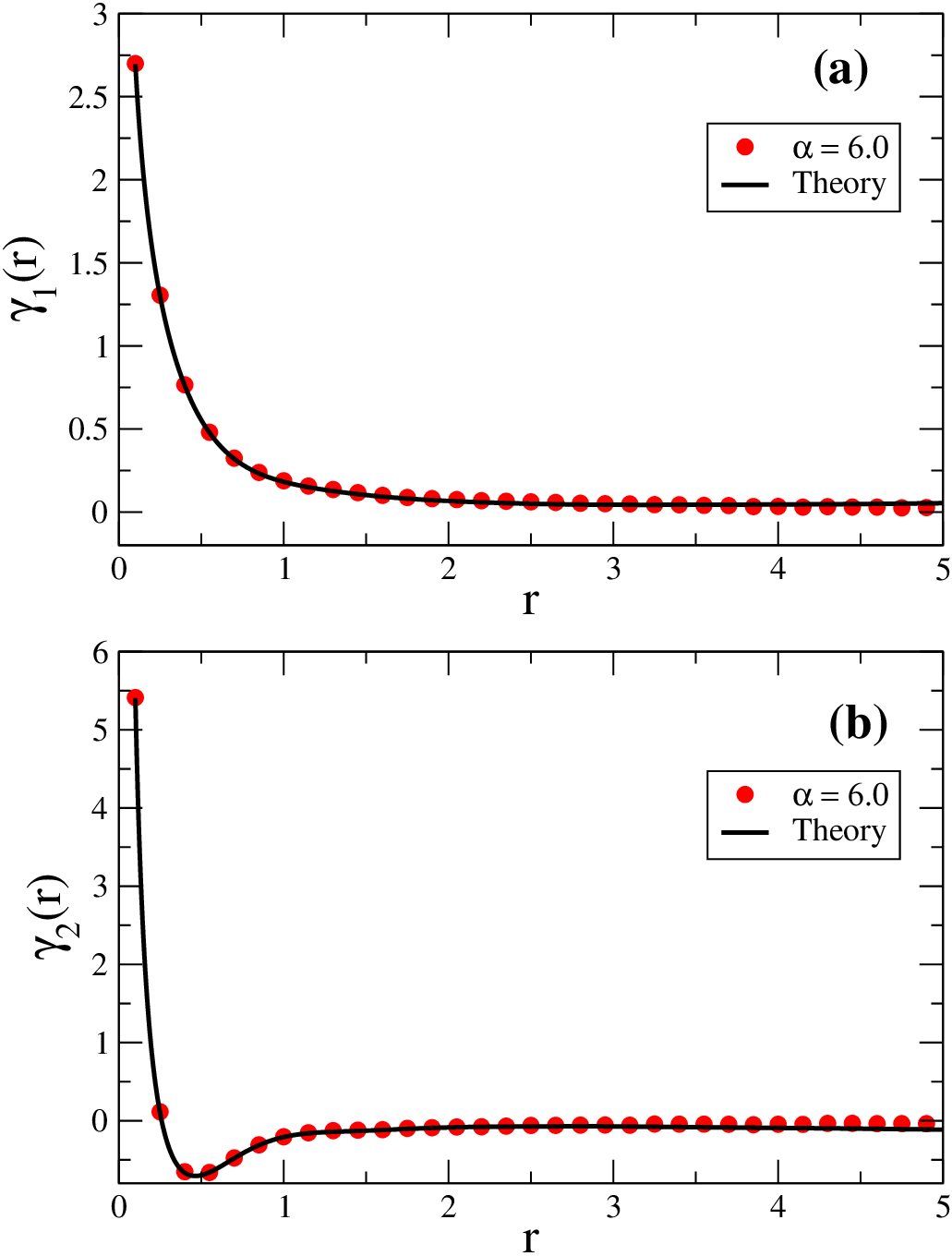}}
\caption{Color Online (a). Skewness parameter of the distribution of $n\left(r\right)$, the number of
levels in intervals of length $r$ shown as a function of $r$ for various values of $\alpha$.
Also shown is the theoretical prediction for the GOE. $N = 4000$ has been used in this calculation.
(b). Excess parameter of the distribution of $n\left(r\right)$, the number of levels
in intervals of length $r$ shown as a function of $r$ for various values of $\alpha$.
Also shown is the theoretical prediction for the GOE. $N = 4000$ has been used in this calculation.}
\label{fig5}
\end{figure}

For unfolding the data, we define $H(\lambda)$ 
to be the number of frequencies equal to or less than $\lambda$ as shown in Fig. 3, 
and $S(\lambda)$ be a smooth function that passes through the staircase function $H$ in the best-fit sense.
In the present case there is no single function that passes smoothly through the whole of $H$
in the best-fit sense. We leave out small regions (about 10\% of the levels) from each end of 
the spectrum and for the remaining levels in the central region marked as I in Fig. 3, we use a quadratic polynomial 
$D(\lambda) = a + b~\lambda + c~\lambda^2$ as an approximation for $S(\lambda)$ over the remaining spectral 
region. It must be stated that there is no particular reason to leave out lower end region of I other than
obtaining better fitting to the data however, it is evident from Fig. 1(b) and Fig. 1(c) that 
towards the upper end of I, more states are localized and hence for the analysis of this paper, 
it is the central regime that we are interested in. The values of $a$, $b$, and $c$ are obtained by a 
standard least-square fitting procedure. We also calculate the misfit function~\cite{p22,p23} corresponding 
to each of the fits to check how well $D(\lambda)$ approximates $S(\lambda)$.  The plot of the misfit function 
we find is qualitatively similar to the plot in Fig. 1(b) of~\cite{p22}. 

In order to reduce the mismatch to lower degree, we eliminate subregions where the mismatch function has 
a very irregular behavior. In the remaining regular subregions we fit quadratic function to the 
misfit function, and we correct $D(\lambda)$ by these quadratic functions to obtain the desired 
unfolding functions. For each of the fluctuation properties reported in this paper, we combine the data 
from all the subregions of all the spectra.

Note that this process of unfolding is quite different from the one used in references~\cite{p21,p22,p23}
in which an exponential function has been to unfold the data followed by the quadratic correction.
In the present case, we could not find any suitable $S(\lambda)$ that could pass through $H$. Hence
we have to resort to this method. However, the advantage of this method of unfolding is that it provides
a suitable way of handling with more complex functions that have no well-defined analytical forms~\cite{p23a}.

The first fluctuation property that we report here is $p(s)$, the distribution of the normalized
nearest-neighbor spacings $s$ of the frequencies of the unfolded spectra.
For this, we first complete the process of unfolding described above for each of the 
spectra using $D(\lambda)$. At this level, we apply the
quadratic correction to this $D(\lambda)$ ignoring small regions where the misfit function
is very irregular. In this way we obtain the unfolded spectra for each of the spectra.
Selection of a random individual spectrum and analysis of each of them separately
indicates that the the spectrum has fluctuation properties associated with the Gaussian orthogonal
ensemble (GOE) of random matrices. To improve statistics, we combine the data for
all the four regions of all the unfolded spectra. This has been plotted in Fig.~4(a) for
three values of $\alpha$ along with the theoretical prediction~\cite{p30,p31,p32,p33}. It can be seen that
the agreement with the theoretical prediction is extremely close, in spite of 
the values of the participation ratios being lower than those found in three dimensions. 

The second fluctuation property we report is the quantity $\Sigma^2\left(r\right)$, the variance of
the number of levels $n\left(r\right)$ within an interval of length $r$ located
randomly in the unfolded spectrum. This is plotted in Fig.~4(b). It must be emphasized that the
quadratic correction applied to the fitting function $D\left(\omega\right)$ is very important
in calculating $\Sigma^2\left(r\right)$. This calculation is extremely sensitive even to very
small errors in the approximation to $S$. The contribution of any
such error to $\Sigma^2\left(r\right)$ grows as $r^2$, whereas the GOE prediction for
$\Sigma^2\left(r\right)$ grows only as ln$\left(r\right)$~\cite{p21,p22,p23}.
Values for $\alpha = 3.5$ almost overlap to the theoretical prediction whereas we do 
see a shift for the other two $\alpha$ values. This may be due to the following possible effects: 
(i) As explained in~\cite{p21,p22,p23}, the exact locations of the irregular 
regions vary in the contour of the misfit function and this might add to the observed shift. A detailed analysis 
using a $\it spectrum-specific$~\cite{p22,p23} choice of subdomains might help in determining the strength 
of this effect on the observed deviation. (ii) As observed from Fig. 1(b), values of participation ratios decrease
with increasing $\alpha$ and this decrease might have a role to play in the observed deviation.  

In Figs.~5(a) and 5(b), we plot the skewness and excess parameters~\cite{p33} of the fluctuations. 
Also included in these plots are the predictions for the GOE. These predictions have been calculated on the
basis of a large ensemble of $500 \times 500$ matrices belonging to the Gaussian Orthogonal Ensemble. Again we 
observe that the agreement with the theoretical prediction is extremely close. Plots show only one value 
of $\alpha$ just for clarity. Othervalues of $\alpha$ also have the same level of agreement to the theory as 
$\alpha=6$.

\section{\bf Conclusions}

Our results  for the DOS of two-dimensional  clusters shows the survival of a two-peak behavior 
when the width of the potential well describing the Morse potential is reduced progressively 
thereby indicating a weak-disorder in contrast to the three-dimensional case. The participation ratios suggest
that the vibrational spectrum has behavior consistent with the extended modes of the system. 
Further, the vibrational spectra has fluctuation properties associated with the GOE of random matrices. For each of the fluctuation properties, agreement with the GOE prediction is extremely close. The observed shifts in the case of the $\Sigma^2\left(r\right)$ calculations suggest that magnitude of the participation ratio might have a role to play in this behavior, however, since calculations of the normalized nearest neighbor spacings distribution and skewness and excess are 
extremely close to the theoretical prediction, it may not be incorrect to say that the system still follows GOE. 

The work leaves us with the following challenges as a part of the future work: (1) The first challenge 
is to generate local minima that have a variable degree of disorder in two-dimensions, at this moment it is 
not clear how to generate such states in two-dimensions that would allow us to do a systematic study with respect to
disorder. (2) What should be the level of disorder before we see a convincing departure in the statistics from GOE. 
The present work shows that the statistics are of GOE type even in a weak-disorder. The reasons for this effect are 
yet to be determined. (3) As mentioned by Hudon et al.~\cite{hudon} in their studies on bulk nanocrystalline materials
that density can influence thermal properties of nanocrystalline materials. Using this argument, it would of great
interest to generate local minima as a function of densities and repeat the calculations for statistics. 

\section{\bf Acknowledgments}
The author would like to thank Peter H. Poole and Subir K. Sarkar for motivation, suggestions,
useful discussions and critical comments on the manuscript; and ACEnet (Canada) for computational resources.

\end{document}